# Working Paper on Organizational Dynamics within Corporate Venture Capital Firms

Michael Rolfes

Alex "Sandy" Pentland

**Abstract**

Corporate venture capital is in the midst of a renaissance. The end of 2015 marked all-time highs both in the number of corporate firms participating in VC deals and in the amount of capital being deployed by corporate VCs[1]. This paper explores, rather than defines, how these firms find success in the wake of this sudden influx of corporate investors. A series of interviews was conducted in order to capture the direct and indirect objectives, philosophies, and modes of operation within some of these corporate VC organizations. During the course of this exploration, numerous operational coherency issues were discovered. Many firms were implicitly incentivizing conflicting and inconsistent behavior among their investment team. Perhaps most surprising, the worst offenders were the more mature corporate VCs who have been in the game for some time. As will be discussed, fundamental evidence suggests that this misalignment is due to lack of attention and commitment at the executive level as corporate strategy evolves.

---

[1] CB Insights. (2016). The 2015 Global Corporate Venture Capital Year in Review. New York, NY: CB Insights.

**Introduction**

Some of today's most prominent incumbent firms – firms like Walmart, Exxon Mobile, Apple, and GM – are worried that their market position will slowly dissipate as they succumb to powerful emerging macro-scale threats such as increasing competitive intensity, shorter windows of opportunity, and the accelerating pace and scale of sociotechnical progress. The traditional counterbalance provided by corporate R&D is weakening[2,3], and so firms stubbornly trudge on, largely producing only incremental innovations in an increasingly dynamic environment[4]. The infusion and adoption of outside innovation is becoming a critical unmet need for incumbent firms within most industries. These firms must invest in, engage with, and integrate outside innovations to sustain (and possibly expand) their market position.

Venture Capital, though traditionally viewed as something of a mercenary business, provides corporate firms a mechanism for infusing outside innovation. It can facilitate long-term collaborative relationships between innovative entrepreneurial firms and more mature incumbent firms. Rather than simply seeking long-shot financial returns, corporate investors direct cash flow to an agile development workforce riding at the forefront of trends and technology, over which they maintain some level of strategic control. Incumbents can thus leverage the nimble, resourceful, and efficient nature of start-ups to inject compelling innovations into their core business lines.

However simply engaging in venture capital and providing funding for start-ups is not enough to provide a sustainable means to protect and expand an incumbent firm's market position. Such an outcome requires thoughtful and deliberate construction, and unwavering commitment from the executive level. The CEO and corporate management team must fully commit and follow through to leverage and utilize their newly available strategic knowledge and resources. Attempted integration of the firms' value networks is essential, and the partnership should be cultivated over the long-term, unless such a partnership proves unproductive and futile. Failure to follow these basic tenets of attention and adoption leave the incumbent bound to and investment likely to produce middling returns at best, handicapped by blurred objectives and incentives.

---

[2] Raelin, J. A., & Balachandra, R. (1985). R&D Project Termination in High-Tech Industries. IEEE Transactions on Engineering Management, Vol. 32, No. 1.
[3] Balachandra, R., & Friar, J. H. (1997). Factors for Success in R&D Projects and New Product Innovation: A Contextual Framework. IEEE Transactions on Engineering Management, Vol. 44, No. 3.
[4] Christensen, C. M. (1997). The Innovator's Dilemma: When New Technologies Cause Great Firms to Fail. Boston, MA: Harvard Business School Press.

**Background and Methods**

While the interview format provided for loose and open qualitative discussion, the conversation was generally focused around a number of central themes in order to capture and aggregate key pieces of information from each of the firms:

1. What was the corporation hoping to accomplish by engaging in VC?
2. How was the corporate VC structured? Who were the key stakeholders, and how were they informed and incentivized?
3. What was the corporate VC's approach and course of action prior-to, during, and after an investment was made?
4. What constituted a "successful" investment?

These themes may appear rather obvious, but the intent was to incite discussion and extract some of the more subtle and unspoken dynamics at play within each organization. In addition, each participant was given basic, high-level knowledge of the research, although the exact hypotheses and ambitions of the study were obscured in an attempt to prevent biasing or leading the respondents. Individual confidentiality and aggregation of results was also assured so that respondents could provide authentic personal responses without consideration of potential consequences or professional conflicts.

Interview participants consisted of core members from 16 corporate venturing organizations, spanning a broad collection of industries. The lifespan and experience of the CVC firms sampled was also diverse, extending from firms less than 1 year old to CVC firms in business for more than a decade. While the sample size demonstrated a fair attempt at polling a representative sample of the corporate venturing community as a whole, a larger sample size certainly would have added further confidence and emphasis to the results of the study.

**Interview Results and Discussion**

There were few clear themes apparent across this set of interviews . There was no standard formula for success, though the particularly successful corporate firms followed their corporate directives quite well. These firms' stated intentions and substantive actions were consistent, and implicit environmental factors were minimized. The other firms, which were the overwhelming majority, had glaring operational coherency issues that impeded their ability to generate even marginal returns. It is these latter firms that became the focus of the analysis.

The majority of the firms interviewed (87%) expressed their primary objective was something other than financial gain. However, half of the participating firms utilized cash returns and IRR to gauge their relative success or failure. Thus, firms with an expressed interest in strategic gain were measuring their investments based upon financial returns. Investors within these organizations were implicitly biased to primarily pursue the most financially attractive deals, and subsequently shoehorn in a strategic justification later. Other empirical studies and cross-industry surveys provided similar evidence of this disjointed relationship. The Global Corporate Venturing organization published a 2016 report revealing that roughly 80% of corporate VC investments were internally evaluated based on IRR[5].

Other factors like compensation, board strategy, and M&A frequency further underscored some of these coherency issues among corporate VC firms. Only half (56%) of interviewed firms stated that their employee compensation structure did not contain any incentives for pursuing financial gain. 25% of firms expressed they took no board seats, regardless of their equity stake, forfeiting rights to valuable information and strategic control of their investment. Roughly 70% of firms indicated that they had no interest in potential acquisition of the start-ups they had funded. In fact, over the past two decades, less than 4% of start-ups based in the US, Europe, and China were acquired by a company whose VC arm had provided earlier funding[6]. It is astounding that incumbents with an existing relationship, symbiotic value networks, and strategic control of a start-up would allow for a potential competitor to acquire that start-up and sever the relationship. This signals lack of commitment and acceptance of the inter-firm partnership by upper management.

---
[5] Global Corporate Venturing. (2016). *The World of Corporate Venturing 2016: The Definitive Guide to the Industry.* London, UK: Mawsonia Ltd.

One of the more subtle operational coherency issues relates to deal syndication among firms. Nearly all of the firms interviewed stated that they frequently partner with other VCs (both institutional and corporate) when making investments. There are numerous reasons to do this, and most come down to sacrificing some strategic control of a particular start-up in exchange for making the investment more financially attractive[7]. For most corporate investors whose motives are primarily strategic, this partnership comes at significant cost. Aside from relinquishing some control and access to the start-up, these corporate investors also enter into a partnership with other investors who may have conflicting objectives, whose value networks may not sufficiently overlap, and whose opinions of what constitutes success may differ. These investors also run the risk of other strategic missteps, such as revealing long-term technological interests to a competitor.

There are many examples of operational inconsistency scattered throughout the compiled interview data. One company conveyed starting their corporate venturing practice had the goal of marketing themselves as being more innovative and forward thinking, but then operated in relative secrecy and declined to disclose any of the start-ups in which they had invested. Another firm indicated a desire to partner with start-ups with the intention of monitoring adjacent markets, but then neglected to take a board seat and effectively relinquished their rights to valuable insights and information. The data was clear; there was an indisputable systemic issue with operational coherence effecting many of the corporate venturing organizations across the industry.

---

[6] Haemmig, M., & Battistini, B. (2015). Corporate Venturing on the Test Bench. London, UK: Mawsonia Ltd.
[7] Brander, J. A., Amit, R., & Antweiler, W. (2002). Venture-Capital Syndication: Improved Venture Selection vs. the Value-Added Hypothesis. Journal of Economics & Management Strategy, Volume 11, Number 3, 423–452.

**Scoring the overall coherence of CVC**

The overall coherence of the participating corporate venturing organizations was rated based upon their responses to the interviews. Each company's internal coherence was rated on a scale from 1 to 4. A score of 4 was awarded to companies whose objectives, actions, and KPIs were fully aligned. A score of 3 was awarded to companies with minor coherency issues, such as a mismatch between their stated objectives and the KPIs used to review performance. A score of 2 was awarded to companies with moderate coherency issues, with a misalignment of multiple factors while still maintaining a mostly consistent overriding objective. These companies might state they are primarily interested in strategic gain, but all other action and assessment is consistent with a company pursuing financial gain. A score of 1 was awarded to companies with major coherency issues, where there was virtually no consistent approach or objective being followed.

Grouping the firms by age, or the number of years they had managed a formal venture capital operation, revealed a direct correlation between firm maturity and the likelihood of diminished operational consistency. Younger corporate VCs, operating for less than 5 years, generally had only minor coherency issues. More mature corporate VCs, operating for 10 years or longer, generally had significant coherency issues. One potential explanation for this phenomenon is the presumption that firm strategy typically drifts over time. Company politics, cultural shifts, staff turnover, management changes, and other factors will continuously adjust and revise a group's present objectives. Without openly and explicitly addressing these strategic changes, operational alignment weakens and attenuates over time. A second potential explanation is that these more mature VCs were among the first in the corporate venturing domain. They were established prior to any real precedent, academic research, or strategic literature being published. Thus, younger firms are advantaged by their ability to leverage this collective knowledge from prior empirical evidence and case study. They are better positioned to properly structure and coordinate their corporate venture capital operation. Full detail of this age-based segmentation is exhibited below in Figure 1.

| Age of CVC | n | Score (mean) | Score (median) |
|---|---|---|---|
| 0-5 years | 5 | 3.0 | 3.0 |
| 5-10 years | 5 | 2.8 | 3.0 |
| 10+ years | 6 | 2.0 | 2.0 |
| __________ | _____ | _____ | _____ |
| Total | 16 | 2.6 | 2.5 |

*Figure 1 – CVC Coherency Ranking Based on Firm Age*

Grouping the firms by industry also revealed some surprising sector-specific trends. Industries facing heavy regulation and standards governance ranked lower than firms in relatively unregulated industries. Corporate venture capital firms operating in relatively unregulated spaces like software, media, and retail boasted higher levels of operational coherence. One potential explanation for this phenomenon is that the barrier to entry is lower in these relatively unregulated industries. It is easier for start-ups to enter, and thus investment opportunities for corporate VCs in these sectors are more plentiful. It is easier for firms to remain focused on their stated objectives, rather than stretching themselves to identify potential deals. It is certainly conceivable that corporate VCs facing fewer investment opportunities would override (or ignore) their firm's objectives in the interest of sustaining their potential deal flow. A second possible explanation stems from the earlier age-related segmentation. Firms in these heavily regulated industries are also generally more mature, and thus more likely to suffer from operational coherency issues. Full detail of this industry-based segmentation is exhibited below in Figure 2.

| Industry of CVC | n | Score (mean) | Score (median) |
|---|---|---|---|
| Software / Enterprise IT | 3 | 3.0 | 3.0 |
| Hardware / Semiconductor | 5 | 2.4 | 2.0 |
| Heavy Industry / Industrial Supply | 3 | 2.3 | 3.0 |
| Chemical / Oil & Gas | 3 | 2.0 | 2.0 |
| Other (Media, Retail) | 2 | 3.5 | 3.5 |
| ___________ | _____ | _____ | _____ |
| Total | 16 | 2.6 | 2.5 |

*Figure 2 – CVC Coherency Ranking Based on Firm Industry*

**Conclusions**

As exhibited below in Figure 3, most corporate venturing organizations' primary objectives, generally established by the CEO or corporate sponsor, were strategic in nature. These firms expressed a desire to establish complementary relationships with smaller entrepreneurial firms in order to maintain their market position. However, numerous organizational and investment practices were in significant discord with these strategic objectives. Each of the considered factors were examined and charted in the plot below. Points farther from the top left of the chart were at greater odds with the majority of firms' stated intentions.

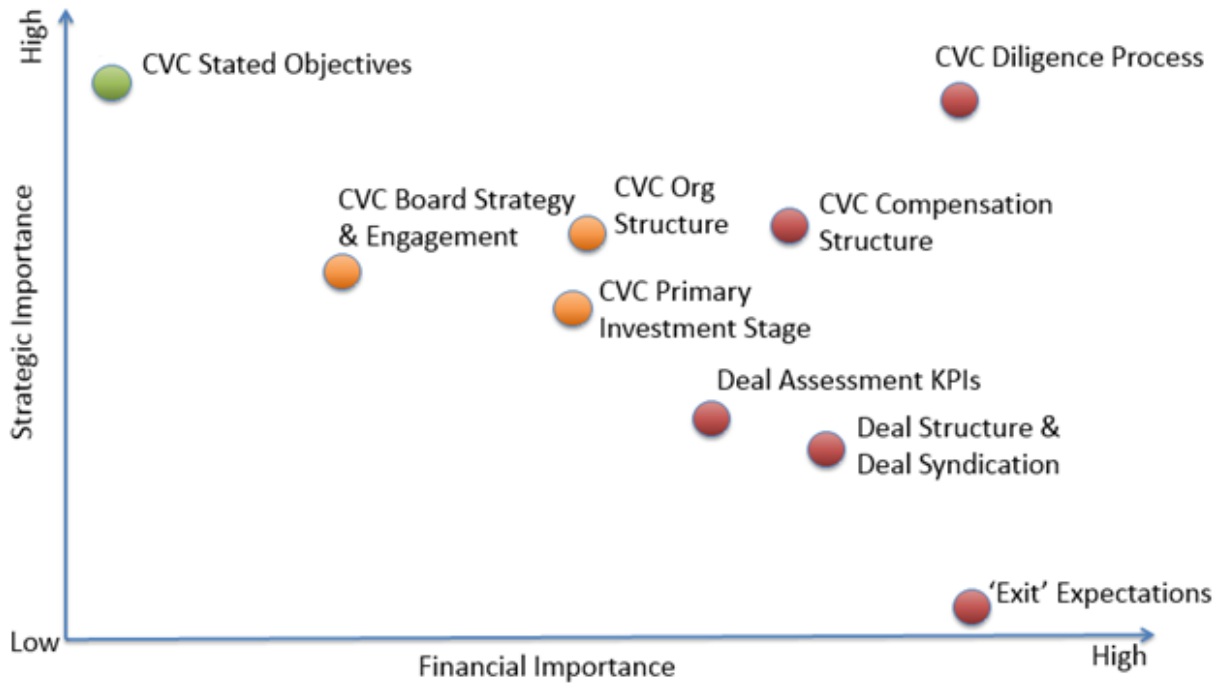

*Figure 3 – Scatter plot of examined factors*

In general, these coherency issues seem to be a result of the corporate executives' lack of focus and resolve to on behalf of the corporate venturing organization. As Michael Porter asserts, "The firm stuck in the middle also probably suffers from a blurred corporate culture and a conflicting set of organizational arrangements and motivation system[8]." The CEO, or corporate sponsor, must establish a cohesive investment strategy that mitigates potential conflicts and implicit biases. If the CEO or corporate sponsor is trying to play both sides of the fence, extracting strategic returns while chasing financial gain, they are almost certainly assured middling outcomes – both financially and strategically. The CEO must establish trust and consistency by structuring, committing and deploying the appropriate resources from across the corporation.

The CEO may fail to commit for a number of a reasons. Liability and legal exposure is heightened for an incumbent when they are involved with a small start-up. Additionally, internal business units and development organizations often have difficulty accepting externally developed innovations. This can slow or inhibit product or platform integration. There is also a degree of uncertainty when committing corporate resources to projects outside the complete control of the organization. CEOs may find it more palatable to sit back and wait for things to develop, rather than take decisive action and assume some level of risk. The start-up and the incumbent may also have intangible issues, such as cultural differences and differing perceptions of time. This can lead to mistrust and potential conflict if not properly addressed early[9][10].

---

[8] Porter, M. (1980). Competitive Strategy. New York, NY: New York Free Press.

The interview responses and other empirical data suggests that many CEOs are succumbing to these diffusive forces. Further, the data signals these CEOs may not be completely focused on the corporate venturing operation, which is even more troubling. Each of the investment teams were given a high-level objective, but were then handicapped by conflicting motivations. Teams instructed to pursue strategic gains were compensated and assessed based on financial returns. Some teams were not even taking board seats, which restricted their entitlement to strategic control and vital information. It is the fault of the executive leadership to have such glaring inconsistencies permeate their VC organization. If not for lack of CEO focus, then perhaps inexperience and lack of commitment can explain these seemingly arbitrary arrangements. One thing is certain here: the CEO needs to get off the sidelines and take a more active role in the corporate venturing operation. The group's ultimate success or failure depends on it.

---